\ifx\pdfoutput\undefined        
  \documentclass[useAMS,usenatbib]{mn2e}
  \usepackage{url}
\else                           
  \documentclass[pdflatex,useAMS,usenatbib]{mn2e}
  \usepackage[bookmarks=false]{hyperref}
\fi

\usepackage{graphicx}
\graphicspath{{./fig/}}
\graphicspath{{./png/}}
\usepackage{bm}
\usepackage{url}

\newcommand{\Div}     {\nabla\cdot}
\newcommand{\uv}      {\bm{u}}
\newcommand{\Laplace} {\nabla^2}
\newcommand{\Strain}  {\bm{\mathsf{S}}}

\title{Mach number dependence of the onset of dynamo action}

\author[Nils Erland L.\ Haugen, Axel Brandenburg, and Antony J.\ Mee]
{Nils Erland L.\ Haugen$^1$, Axel Brandenburg$^2$, and Antony J.\ Mee$^3$\\
$^{1}$Department of Physics, The Norwegian University of Science
  and Technology, H{\o}yskoleringen 5, N-7034 Trondheim, Norway\\
$^2$NORDITA, Blegdamsvej 17, DK-2100 Copenhagen \O, Denmark\\
$^3$School of Mathematics and Statistics, University of Newcastle,
  Newcastle upon Tyne, NE1 7RU, UK}

\begin{document}


\newcommand{\EQ}{\begin{equation}}
\newcommand{\EN}{\end{equation}}
\newcommand{\EQA}{\begin{eqnarray}}
\newcommand{\ENA}{\end{eqnarray}}
\newcommand{\eq}[1]{(\ref{#1})}
\newcommand{\EEq}[1]{Equation~(\ref{#1})}
\newcommand{\Eq}[1]{equation~(\ref{#1})}
\newcommand{\Eqs}[2]{equations~(\ref{#1}) and~(\ref{#2})}
\newcommand{\Eqss}[2]{equations~(\ref{#1})--(\ref{#2})}
\newcommand{\eqs}[2]{(\ref{#1}) and~(\ref{#2})}
\newcommand{\App}[1]{Appendix~\ref{#1}}
\newcommand{\Sec}[1]{Sect.~\ref{#1}}
\newcommand{\Secs}[2]{\S\S\ref{#1} and \ref{#2}}
\newcommand{\Fig}[1]{Fig.~\ref{#1}}
\newcommand{\FFig}[1]{Figure~\ref{#1}}
\newcommand{\Tab}[1]{Table~\ref{#1}}
\newcommand{\Figs}[2]{Figs~\ref{#1} and \ref{#2}}
\newcommand{\Tabs}[2]{Tables~\ref{#1} and \ref{#2}}
\newcommand{\bra}[1]{\langle #1\rangle}
\newcommand{\bbra}[1]{\left\langle #1\right\rangle}
\newcommand{\mean}[1]{\overline #1}
\newcommand{\meanemf}{\overline{\mbox{\boldmath ${\cal E}$}}{}}{}
\newcommand{\meanemfs}{\overline{\cal E} {}}
\newcommand{\meanAA}{\overline{\bf{A}}}
\newcommand{\meanBB}{\overline{\bf{B}}}
\newcommand{\meanJJ}{\overline{\bf{J}}}
\newcommand{\meanUU}{\overline{\bf{U}}}
\newcommand{\meanuu}{\overline{\mbox{\boldmath $u$}}{}}{}
\newcommand{\meanoo}{\overline{\mbox{\boldmath $\omega$}}{}}{}
\newcommand{\meanEE}{\overline{\mbox{\boldmath ${\cal E}$}}{}}{}
\newcommand{\meanuxB}{\overline{\mbox{\boldmath $\delta u\times \delta B$}}{}}{}
\newcommand{\meanJB}{\overline{\mbox{\boldmath $J\cdot B$}}{}}{}
\newcommand{\meanAB}{\overline{\mbox{\boldmath $A\cdot B$}}{}}{}
\newcommand{\meanjb}{\overline{\mbox{\boldmath $j\cdot b$}}{}}{}
\newcommand{\meanA}{\overline{A}}
\newcommand{\meanB}{\overline{B}}
\newcommand{\meanC}{\overline{C}}
\newcommand{\meanU}{\overline{U}}
\newcommand{\meanJ}{\overline{J}}
\newcommand{\meanS}{\overline{S}}
\newcommand{\meanF}{\overline{\cal F}}
%
%
\newcommand{\teps}{\tilde{\epsilon} {}}
\newcommand{\zh}{\hat{z}}
%
%
\newcommand{\eee}{\hat{\mbox{\boldmath $e$}} {}}
\newcommand{\nnn}{\hat{\mbox{\boldmath $n$}} {}}
\newcommand{\vvv}{\hat{\mbox{\boldmath $v$}} {}}
\newcommand{\rr}{\hat{\mbox{\boldmath $r$}} {}}
\newcommand{\xxx}{\hat{\bf x}}
\newcommand{\yyy}{\hat{\bf y}}
\newcommand{\zz}{\hat{\bf z}}
\newcommand{\pphi}{\hat{\bm\phi}}
\newcommand{\ttt}{\hat{\mbox{\boldmath $\theta$}} {}}
\newcommand{\OOO}{\hat{\mbox{\boldmath $\Omega$}} {}}
\newcommand{\ooo}{\hat{\mbox{\boldmath $\omega$}} {}}
\newcommand{\BBBB}{\hat{\mbox{\boldmath $B$}} {}}
\newcommand{\Bhat}{\hat{B}}
\newcommand{\BBhat}{\hat{\bm{B}}}
%
%
\newcommand{\gggg}{\mbox{\boldmath $g$} {}}
\newcommand{\ddd}{\mbox{\boldmath $d$} {}}
\newcommand{\rrr}{\mbox{\boldmath $r$} {}}
\newcommand{\yy}{\mbox{\boldmath $y$} {}}
\newcommand{\zzz}{\mbox{\boldmath $z$} {}}
\newcommand{\vv}{\mbox{\boldmath $v$} {}}
\newcommand{\ww}{\mbox{\boldmath $w$} {}}
\newcommand{\mm}{\mbox{\boldmath $m$} {}}
\newcommand{\PP}{\mbox{\boldmath $P$} {}}
\newcommand{\QQ}{\mbox{\boldmath $Q$} {}}
\newcommand{\bp}{\mbox{\boldmath $p$} {}}
\newcommand{\pp}{\mbox{\boldmath $p$} {}}
\newcommand{\qq}{\mbox{\boldmath $q$} {}}
\newcommand{\II}{\mbox{\boldmath $I$} {}}
\newcommand{\xx}{{\bm{x}}}
\newcommand{\UU}{{\bm{U}}}
\newcommand{\uu}{{\bm{u}}}
\newcommand{\BB}{{\bm{B}}}
\newcommand{\HH}{{\bm{H}}}
\newcommand{\CC}{{\bm{C}}}
\newcommand{\JJ}{{\bm{J}}}
\newcommand{\jj}{{\bm{j}}}
\newcommand{\AAA}{{\bm{A}}}
\newcommand{\aaaa}{{\bm{a}}}
\newcommand{\bb}{{\bm{b}}}
\newcommand{\cc}{{\bm{c}}}
\newcommand{\ee}{{\bm{e}}}
\newcommand{\nn}{\mbox{\boldmath $n$} {}}
\newcommand{\ff}{\mbox{\boldmath $f$} {}}
\newcommand{\hh}{\mbox{\boldmath $h$} {}}
\newcommand{\EE}{{\bm{E}}}
\newcommand{\FF}{{\bm{F}}}
\newcommand{\KK}{{\bm{K}}}
\newcommand{\kk}{{\bm{k}}}
\newcommand{\TT}{\mbox{\boldmath $T$} {}}
\newcommand{\MM}{\mbox{\boldmath $M$} {}}
\newcommand{\GG}{\mbox{\boldmath $G$} {}}
\newcommand{\SSS}{{\mf{S}}}
\newcommand{\grav}{\mbox{\boldmath $g$} {}}
\newcommand{\nab}{\mbox{\boldmath $\nabla$} {}}
\newcommand{\OO}{\mbox{\boldmath $\Omega$} {}}
\newcommand{\oo}{\mbox{\boldmath $\omega$} {}}
\newcommand{\ttau}{\mbox{\boldmath $\tau$} {}}
\newcommand{\LL}{\mbox{\boldmath $\Lambda$} {}}
\newcommand{\llambda}{\mbox{\boldmath $\lambda$} {}}
\newcommand{\pomega}{\mbox{\boldmath $\varpi$} {}}
%
%
\newcommand{\SSSS}{\mbox{\boldmath ${\sf S}$} {}}
\newcommand{\LLLL}{\mbox{\boldmath ${\sf L}$} {}}
\newcommand{\PPPP}{\mbox{\boldmath ${\sf P}$} {}}
\newcommand{\MMMM}{\mbox{\boldmath ${\sf M}$} {}}
\newcommand{\AAAA}{\mbox{\boldmath ${\cal A}$} {}}
\newcommand{\BBB}{\mbox{\boldmath ${\cal B}$} {}}
\newcommand{\emf}{\mbox{\boldmath ${\cal E}$} {}}
\newcommand{\FFF}{\mbox{\boldmath ${\cal F}$} {}}
\newcommand{\GGG}{\mbox{\boldmath ${\cal G}$} {}}
\newcommand{\HHH}{\mbox{\boldmath ${\cal H}$} {}}
\newcommand{\QQQ}{\mbox{\boldmath ${\cal Q}$} {}}
\newcommand{\GGGG}{{\bf G} {}}
%
%
\newcommand{\ii}{{\rm i}}
\newcommand{\grad}{{\rm grad} \, {}}
\newcommand{\curl}{{\rm curl} \, {}}
\newcommand{\dive}{{\rm div}  \, {}}
\newcommand{\Dive}{{\rm Div}  \, {}}
\newcommand{\sgn}{{\rm sgn}  \, {}}
\newcommand{\DD}{{\rm D} {}}
\newcommand{\DDD}{{\cal D} {}}
\newcommand{\dd}{{\rm d} {}}
\newcommand{\const}{{\rm const}  {}}
\newcommand{\CR}{{\rm CR}}
\def\degr{\hbox{$^\circ$}}
\def\la{\mathrel{\mathchoice {\vcenter{\offinterlineskip\halign{\hfil
$\displaystyle##$\hfil\cr<\cr\sim\cr}}}
{\vcenter{\offinterlineskip\halign{\hfil$\textstyle##$\hfil\cr<\cr\sim\cr}}}
{\vcenter{\offinterlineskip\halign{\hfil$\scriptstyle##$\hfil\cr<\cr\sim\cr}}}
{\vcenter{\offinterlineskip\halign{\hfil$\scriptscriptstyle##$\hfil\cr<\cr\sim\cr}}}}}
\def\ga{\mathrel{\mathchoice {\vcenter{\offinterlineskip\halign{\hfil
$\displaystyle##$\hfil\cr>\cr\sim\cr}}}
{\vcenter{\offinterlineskip\halign{\hfil$\textstyle##$\hfil\cr>\cr\sim\cr}}}
{\vcenter{\offinterlineskip\halign{\hfil$\scriptstyle##$\hfil\cr>\cr\sim\cr}}}
{\vcenter{\offinterlineskip\halign{\hfil$\scriptscriptstyle##$\hfil\cr>\cr\sim\cr}}}}}
%
%
\def\Ta{\mbox{\rm Ta}}
\def\Ra{\mbox{\rm Ra}}
\def\Ma{\mbox{\rm Ma}}
\def\Co{\mbox{\rm Co}}
\def\Roo{\mbox{\rm Ro}^{-1}}
\def\Rooo{\mbox{\rm Ro}^{-2}}
\def\Pra{\mbox{\rm Pr}}
\def\Pran{\mbox{\rm Pr}}
\def\Pm{\mbox{\rm Pr}_M}
\def\Rm{\mbox{\rm Re}_M}
\def\Rey{\mbox{\rm Re}}
\def\Pe{\mbox{\rm Pe}}
\newcommand{\ea}{{\rm et al.\ }}
\newcommand{\eaa}{{\rm et al.\ }}
\def\half{{\textstyle{1\over2}}}
\def\threehalf{{\textstyle{3\over2}}}
\def\onethird{{\textstyle{1\over3}}}
\def\onesixth{{\textstyle{1\over6}}}
\def\twothird{{\textstyle{2\over3}}}
\def\fourthird{{\textstyle{4\over3}}}
\def\quarter{{\textstyle{1\over4}}}
\newcommand{\W}{\,{\rm W}}
\newcommand{\V}{\,{\rm V}}
\newcommand{\kV}{\,{\rm kV}}
\newcommand{\T}{\,{\rm T}}
\newcommand{\G}{\,{\rm G}}
\newcommand{\Hz}{\,{\rm Hz}}
\newcommand{\kHz}{\,{\rm kHz}}
\newcommand{\kG}{\,{\rm kG}}
\newcommand{\K}{\,{\rm K}}
\newcommand{\g}{\,{\rm g}}
\newcommand{\s}{\,{\rm s}}
\newcommand{\ms}{\,{\rm ms}}
\newcommand{\ks}{\,{\rm ks}}
\newcommand{\cm}{\,{\rm cm}}
\newcommand{\m}{\,{\rm m}}
\newcommand{\km}{\,{\rm km}}
\newcommand{\kms}{\,{\rm km/s}}
\newcommand{\kg}{\,{\rm kg}}
\newcommand{\ug}{\,\mu{\rm g}}
\newcommand{\kW}{\,{\rm kW}}
\newcommand{\MW}{\,{\rm MW}}
\newcommand{\Mm}{\,{\rm Mm}}
\newcommand{\Mx}{\,{\rm Mx}}
\newcommand{\pc}{\,{\rm pc}}
\newcommand{\kpc}{\,{\rm kpc}}
\newcommand{\yr}{\,{\rm yr}}
\newcommand{\Myr}{\,{\rm Myr}}
\newcommand{\Gyr}{\,{\rm Gyr}}
\newcommand{\erg}{\,{\rm erg}}
\newcommand{\mol}{\,{\rm mol}}
\newcommand{\dyn}{\,{\rm dyn}}
\newcommand{\J}{\,{\rm J}}
\newcommand{\RM}{\,{\rm RM}}
\newcommand{\EM}{\,{\rm EM}}
\newcommand{\AU}{\,{\rm AU}}
\newcommand{\A}{\,{\rm A}}
%
%
\newcommand{\yastroph}[2]{ #1, astro-ph/#2}
\newcommand{\ycsf}[3]{ #1, {Chaos, Solitons \& Fractals,} {#2}, #3}
\newcommand{\yepl}[3]{ #1, {Europhys. Lett.,} {#2}, #3}
\newcommand{\yaj}[3]{ #1, {AJ,} {#2}, #3}
\newcommand{\yjgr}[3]{ #1, {JGR,} {#2}, #3}
\newcommand{\ysol}[3]{ #1, {Sol. Phys.,} {#2}, #3}
\newcommand{\yapj}[3]{ #1, {ApJ,} {#2}, #3}
\newcommand{\yapjl}[3]{ #1, {ApJ,} {#2}, #3}
\newcommand{\yapjs}[3]{ #1, {ApJ Suppl.,} {#2}, #3}
\newcommand{\yan}[3]{ #1, {AN,} {#2}, #3}
\newcommand{\ymhdn}[3]{ #1, {Magnetohydrodyn.} {#2}, #3}
\newcommand{\yana}[3]{ #1, {A\&A,} {#2}, #3}
\newcommand{\yanas}[3]{ #1, {A\&AS,} {#2}, #3}
\newcommand{\yanar}[3]{ #1, {A\&AR,} {#2}, #3}
\newcommand{\yass}[3]{ #1, {Ap\&SS,} {#2}, #3}
\newcommand{\ygafd}[3]{ #1, {Geophys. Astrophys. Fluid Dyn.,} {#2}, #3}
\newcommand{\ypasj}[3]{ #1, {Publ. Astron. Soc. Japan,} {#2}, #3}
\newcommand{\yjfm}[3]{ #1, {JFM,} {#2}, #3}
\newcommand{\ypf}[3]{ #1, {Phys. Fluids,} {#2}, #3}
\newcommand{\ypp}[3]{ #1, {Phys. Plasmas,} {#2}, #3}
\newcommand{\ysov}[3]{ #1, {Sov. Astron.,} {#2}, #3}
\newcommand{\ysovl}[3]{ #1, {Sov. Astron. Lett.,} {#2}, #3}
\newcommand{\yjetp}[3]{ #1, {Sov. Phys. JETP,} {#2}, #3}
\newcommand{\yphy}[3]{ #1, {Physica,} {#2}, #3}
\newcommand{\yannr}[3]{ #1, {ARA\&A,} {#2}, #3}
\newcommand{\yaraa}[3]{ #1, {ARA\&A,} {#2}, #3}
\newcommand{\yprs}[3]{ #1, {Proc. Roy. Soc. Lond.,} {#2}, #3}
\newcommand{\yprl}[3]{ #1, {PRL,} {#2}, #3}
\newcommand{\ypre}[3]{ #1, {PRE,} {#2}, #3}
\newcommand{\yphl}[3]{ #1, {Phys. Lett.,} {#2}, #3}
\newcommand{\yptrs}[3]{ #1, {Phil. Trans. Roy. Soc.,} {#2}, #3}
\newcommand{\ymn}[3]{ #1, {MNRAS,} {#2}, #3}
\newcommand{\ynat}[3]{ #1, {Nat,} {#2}, #3}
\newcommand{\ysci}[3]{ #1, {Sci,} {#2}, #3}
\newcommand{\ysph}[3]{ #1, {Solar Phys.,} {#2}, #3}
\newcommand{\ypr}[3]{ #1, {Phys. Rev.,} {#2}, #3}
\newcommand{\spr}[2]{ ~#1~ {\em Phys. Rev. }{\bf #2} (submitted)}
\newcommand{\ppr}[2]{ ~#1~ {\em Phys. Rev. }{\bf #2} (in press)}
\newcommand{\ypnas}[3]{ #1, {Proc. Nat. Acad. Sci.,} {#2}, #3}
\newcommand{\yicarus}[3]{ #1, {Icarus,} {#2}, #3}
\newcommand{\yspd}[3]{ #1, {Sov. Phys. Dokl.,} {#2}, #3}
\newcommand{\yjcp}[3]{ #1, {J. Comput. Phys.,} {#2}, #3}
\newcommand{\yjour}[4]{ #1, {#2}, {#3}, #4}
\newcommand{\yprep}[2]{ #1, {\sf #2}}
\newcommand{\ybook}[3]{ #1, {#2} (#3)}
\newcommand{\yproc}[5]{ #1, in {#3}, ed. #4 (#5), #2}
\newcommand{\pproc}[4]{ #1, in {#2}, ed. #3 (#4), (in press)}
\newcommand{\ppp}[1]{ #1, {Phys. Plasmas,} (in press)}
\newcommand{\sapj}[1]{ #1, {ApJ,} (submitted)}
\newcommand{\sana}[1]{ #1, {A\&A,} (submitted)}
\newcommand{\san}[1]{ #1, {AN,} (submitted)}
\newcommand{\sprl}[1]{ #1, {PRL,} (submitted)}
\newcommand{\pprl}[1]{ #1, {PRL,} (in press)}
\newcommand{\sjfm}[1]{ #1, {JFM,} (submitted)}
\newcommand{\sgafd}[1]{ #1, {Geophys. Astrophys. Fluid Dyn.,} (submitted)}
\newcommand{\pgafd}[1]{ #1, {Geophys. Astrophys. Fluid Dyn.,} (in press)}
\newcommand{\tana}[1]{ #1, {A\&A,} (to be submitted)}
\newcommand{\smn}[1]{ #1, {MNRAS,} (submitted)}
\newcommand{\pmn}[1]{ #1, {MNRAS,} (in press)}
\newcommand{\papj}[1]{ #1, {ApJ,} (in press)}
\newcommand{\papjl}[1]{ #1, {ApJL,} (in press)}
\newcommand{\sapjl}[1]{ #1, {ApJL,} (submitted)}
\newcommand{\pana}[1]{ #1, {A\&A,} (in press)}
\newcommand{\pan}[1]{ #1, {AN,} (in press)}
\newcommand{\pjour}[2]{ #1, {#2,} (in press)}
\newcommand{\sjour}[2]{ #1, {#2,} (submitted)}

\date{\today,~ $ $Revision: 1.106 $ $}

\pagerange{\pageref{firstpage}--\pageref{lastpage}} 

\maketitle

\label{firstpage}

\begin{abstract}
The effect of compressibility on the onset of nonhelical turbulent
dynamo action is investigated using both direct simulations as well as
simulations with shock-capturing viscosities, keeping however the
regular magnetic diffusivity.
It is found that the critical magnetic Reynolds number increases from
about 35 in the subsonic regime to about 70 in the supersonic regime.
Although the shock structures are sharper in the high resolution direct
simulations compared to the low resolution shock-capturing simulations,
the magnetic field looks roughly similar in both cases
and does not show shock structures.
Similarly, the onset of dynamo action is not significantly
affected by the shock-capturing viscosity.
\end{abstract}

\section{Introduction}

Transonic and supersonic turbulence is widespread in many astrophysical
settings where magnetic fields are believed to be generated and maintained
by dynamo action.
Examples include supernova-driven turbulence in the interstellar medium
\citep{Korpi99,deAvillezMacLow02,Bals04} and turbulence in galaxy clusters
\citep{Roettiger99} where the driving comes mainly from
cluster mergers.
In these cases the root mean square (rms) Mach number is of the order of
unity, but in the cooler parts of the interstellar medium the sound
speed is low and the flows can therefore easily become highly
supersonic and may reach rms Mach numbers of around 20
\citep{Padoan97,Padoan02}.

Simulations do not give a clear picture of how the excitation conditions
for dynamo action change in the highly supersonic regime.
While there are clear examples of dynamo action in supersonic turbulence
\citep{Bals04}, the dynamo seems to be less efficient in the more strongly
supersonic regime \citep{Padoan04}.
It is plausible that newly generated magnetic field gets too quickly
entrained by the shocks where the field is then dissipated \citep{Padoan99}.

Supersonic turbulence has a significant irrotational component.
Purely irrotational turbulence is also referred to as acoustic turbulence
and can be described using weak turbulence theory \citep{ZS70,Byk88}.
Since supersonic turbulence contains shocks, such flows can also be
described as shock turbulence \citep{KP73}.
If these flows were described as predominantly irrotational,
the growth rate would
increase with the Mach number to the fourth power \citep{KRS85,MS96}.
However, this tendency has never been seen in simulations.

There are however caveats to both the analytical and the numerical
approaches.
Firstly, it is clear that even highly supersonic turbulence is not
fully irrotational and that still $70$-$80\%$ of the kinetic energy
comes from the solenoidal component \citep{Porter98,Padoan99}.
Therefore purely acoustic turbulence cannot be used as an
approximation to supersonic turbulence.
The other problem that we shall be concerned with here is the question
to what extent does the numerical shock viscosity used in many simulations
affect the conclusion regarding dynamo action.

In order to address the latter question we perform direct simulations at
sufficiently high resolution so that no numerical shock viscosity
is needed.
Here, the term `direct' means that one uses the microscopic viscosity,
i.e.\ one ignores the fact that in many astrophysical applications one
will never be able to reach realistic Reynolds numbers.
We also compare with calculations where a shock-capturing viscosity
is included.
Here the viscosity is locally enhanced in a shock, allowing less
diffusion between the shocks.
It turns out that both direct and shock-capturing viscosity
simulations predict an approximately similar increase in the
critical magnetic Reynolds number for dynamo action.
In these cases the magnetic diffusivity is the same
as in the direct simulations.

\section{Method}

We consider an isothermal gas with constant sound speed $c_{\rm s}$.
The continuity equation is
written in terms of the logarithmic density,
\EQ
{\DD\ln\rho\over\DD t}=-\nab\cdot\uu,
\EN
and the induction equation is solved in terms of the magnetic vector
potential $\AAA$, where $\BB=\nab\times\AAA$, and
\EQ
{\partial\AAA\over\partial t}=\uu\times\BB+\eta\nabla^2\AAA,
\label{dAdt}
\EN
$\eta$ being the magnetic diffusivity.
The momentum equation is solved in the form
\EQ
{\DD\uu\over\DD t}=-c_{\rm s}^2\nab\ln\rho+\rho^{-1}\left(\JJ\times\BB
+\FF_{\rm visc}+\ff\right),
\label{dudt}
\EN
where $\DD/\DD t=\partial/\partial t+\uu\cdot\nab$ is the advective
derivative, $\JJ=\nab\times\BB/\mu_0$ the current density,
$\BB$ the magnetic field, $\mu_0$
the vacuum permeability,
$\FF_{\rm visc}$ is the viscous force (see below),
and $\ff$ is a random forcing function with
\EQ
\ff(\xx,t)=\Rey\{N\ff_{\kk(t)}\exp[\ii\kk(t)\cdot\xx+\ii\phi(t)]\},
\EN
where $\xx$ is the position vector.
The wavevector $\kk(t)$ and the random phase
$-\pi<\phi(t)\le\pi$ change at every time step, so $\ff(\xx,t)$ is,
up to discretization errors, $\delta$-correlated in time.
We force the system with nonhelical transversal waves,
\EQ
\ff_{\kk}=\left(\kk\times\eee\right)/\sqrt{\kk^2-(\kk\cdot\eee)^2},
\label{nohel_forcing}
\EN
where $\eee$ is an arbitrary unit vector not aligned with $\kk$;
note that $|\ff_{\kk}|^2=1$.
On dimensional grounds the normalization factor $N$ is chosen to be
$N=f_0 c_{\rm s}(|\kk|c_{\rm s}/\delta t)^{1/2}$, where $f_0$ is a
nondimensional forcing amplitude.
We use the \textsc{Pencil Code},\footnote{
\url{http://www.nordita.dk/data/brandenb/pencil-code}}
which is a high-order
finite-difference code (sixth order in space and third
order in time) for solving the compressible hydromagnetic equations.

In the direct simulations with constant viscosity, $\nu$, the viscous
force per unit mass is given by
\begin{equation}
\bm{F}_{\rm visc}^{(\nu)}
=\rho\nu
\left(\Laplace\uv+\onethird\nab\nab\cdot\uv
+2\Strain\cdot\nab\ln\rho\right),
\end{equation}
where $\rho$ is the density and
\begin{equation}
{\sf S}_{ij}=
\frac{1}{2}\left({\partial u_i\over\partial x_j}
+ {\partial u_j\over\partial x_i}\right)
-\frac{1}{3}\delta_{ij}\nab\cdot\uu
\end{equation}
is the traceless rate of strain matrix.
In runs with shock-capturing viscosity we simply add to this a spatially
dependent diffusion term such that the effective viscosity is enhanced
only in the neighbourhood of a shock \citep{RichtmyerMorton67}.
This technique artificially broadens the shocks such that
they can be resolved numerically, and hence all the conservation
laws are obeyed, which is important for satisfying the right jump
conditions.
It is sufficient to broaden the shocks using only a
bulk viscosity, $\zeta$, rather than a locally enhanced shear viscosity.
Thus we write
\begin{equation}
\bm{F}_{\rm visc}^{\rm(shock)}
=\bm{F}_{\rm visc}^{(\nu)}
+\rho\zeta\nab\nab\cdot\uv
+(\nab\cdot\uv)\nab(\rho\zeta),
\end{equation}
so the full stress tensor is given by
\begin{equation}
\bm{\tau}_{ij}=2\rho\nu{\sf S}_{ij}+\rho\zeta\delta_{ij}\nab\cdot\uv.
\end{equation}
Here $\zeta$ is the shock viscosity.
Following \cite{NG95}, we assume $\zeta$ to be
proportional to the smoothed (over 3 zones)
maximum (over 5 zones) of the positive part of the negative
divergence of the velocity, i.e.\
\begin{equation}
\zeta=c_{\rm shock}\left<\max_5[(-\Div\uv)_+]\right>,
\end{equation}
where $c_{\rm shock}$ is a constant defining the strength of the
shock viscosity.
This is also the technique used by \cite{Padoan02}.

The simulations are governed by three important parameters,
the Mach number $\Ma=u_{\rm rms}/c_{\rm s}$,
the Reynolds number $\Rey=u_{\rm rms}/(\nu k_{\rm f})$,
and the magnetic Prandtl number $\Pm=\nu/\eta$.
The magnetic Reynolds number is defined as $\Rm=\Rey\,\Pm$,
and the critical magnetic Reynolds number is the value 
above which a weak seed magnetic field grows exponentially.

The Mach number can be increased either by increasing the strength
of the forcing or by lowering the sound speed.
We choose the former and keep the sound speed constant.
Moreover, we use $c_{\rm s}$ as our velocity unit.
The lowest wavenumber in the box, $k_1=2\pi/L$, is used as
our inverse length unit.
This implies that time is measured in units of $(c_{\rm s}k_1)^{-1}$.
Density is measured in units of the mean density, which is also equal
to the initial density $\rho_0$.
The extent of the box is in each direction $L=2\pi$,
so the smallest possible wavenumber 
in the box is therefore $k=1$.
The maximum possible wavenumber depends on the resolution and is
$k=256$ for our largest run with $512^3$ meshpoints.
Larger resolution is technically already possible \citep{HBD03},
but still too demanding in terms of computing time
if one wants to cover many turnover times.
In all our simulations
the flow is forced in a band of wavenumbers between 1 and 2.
As initial conditions we have used zero velocity and a weak random
magnetic field with $\bra{\BB^2}/(\mu_0\rho_0 c_{\rm s}^2)\approx10^{-8}$.
We use periodic boundaries in all three directions.  

\section{Results}
\label{SResults}

In \Tab{Trunscompare} we give the Mach and Reynolds numbers as well
as the growth rate for runs with three different combinations of $f_0$
and $\nu$, each with and without added shock viscosity.
The run parameters are generally chosen such that the growth rates
are close to zero so that the critical magnetic Reynolds number for
dynamo action can accurately be determined via interpolation.
It turns out that the addition of shock viscosity reduces the
{\it normalized} growth rate of the dynamo, $\lambda/\lambda_0$,
only by about 0.003--0.016 (see \Tab{Trunscompare}),
which is about 3--16\% of the typical normalized growth
rates of about 0.1 in the more strongly supercritical cases \cite[see
Fig.~3 of][]{HBD04}.
Here, $\lambda_0\equiv u_{\rm rms}k_{\rm f})$ is the typical stretching rate.
The reduction of the growth rates can,
at least partially, be explained by a reduction in 
$\Rey_{\rm M}$ due to reduced rms velocity when shock viscosity is added.

\begin{table}
\caption{
The Mach and Reynolds numbers, as well as the growth rate
(in units of $\lambda_0\equiv u_{\rm rms}k_{\rm f}$)
for runs with different forcing strength ($f_0$),
viscosity ($\nu$) and $c_{\rm shock}$, and number of meshpoints $N$.
All runs have $\Pm=1$.
The runs with only direct viscosity are Runs~1a-3a.
\label{Trunscompare}}
\begin{tabular}{cccccccc}
Run& $N$ & $f_0$& $\nu$&$\Ma$&$\Rey$& $\lambda/\lambda_0$ & $c_{\rm shock}$\\
\hline                                                                 
1a & 512 & 0.2 & 0.006 & 0.72 &  78 &$+0.030$& -\\ 
1b & 128 & 0.2 & 0.006 & 0.70 &  78 &$+0.010$& 0.8\\ 
1c & 128 & 0.2 & 0.006 & 0.68 &  74 &$+0.014$& 3\\ 
\hline                                                                 
2a & 256 & 0.2 & 0.01  & 0.65 &  43 &$+0.006$& -\\ 
2b &  64 & 0.2 & 0.01  & 0.62 &  41 &$-0.004$& 0.8\\ 
2c &  64 & 0.2 & 0.01  & 0.65 &  43 &$-0.005$& 3\\ 
\hline                                                                 
3a & 512 & 0.5 & 0.01  & 1.11 &  75 &$+0.005$& -\\ 
3b & 128 & 0.5 & 0.01  & 1.14 &  76 &$+0.010$& 0.8\\ 
3c &  64 & 0.5 & 0.01  & 1.10 &  74 &$+0.002$& 3\\ 
\end{tabular}
\end{table}

\begin{figure}\centering
\includegraphics[width=0.46\textwidth]{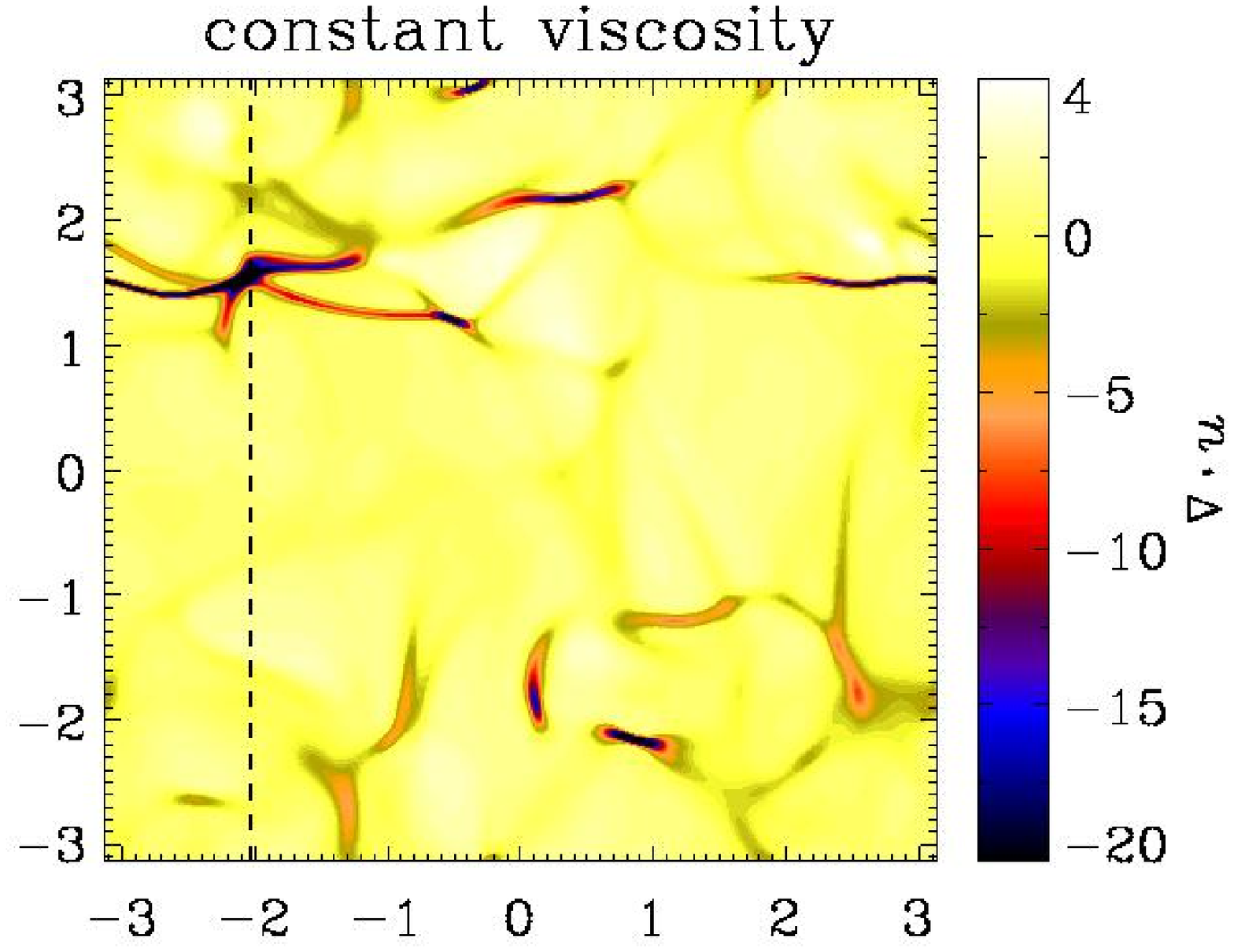}
\includegraphics[width=0.46\textwidth]{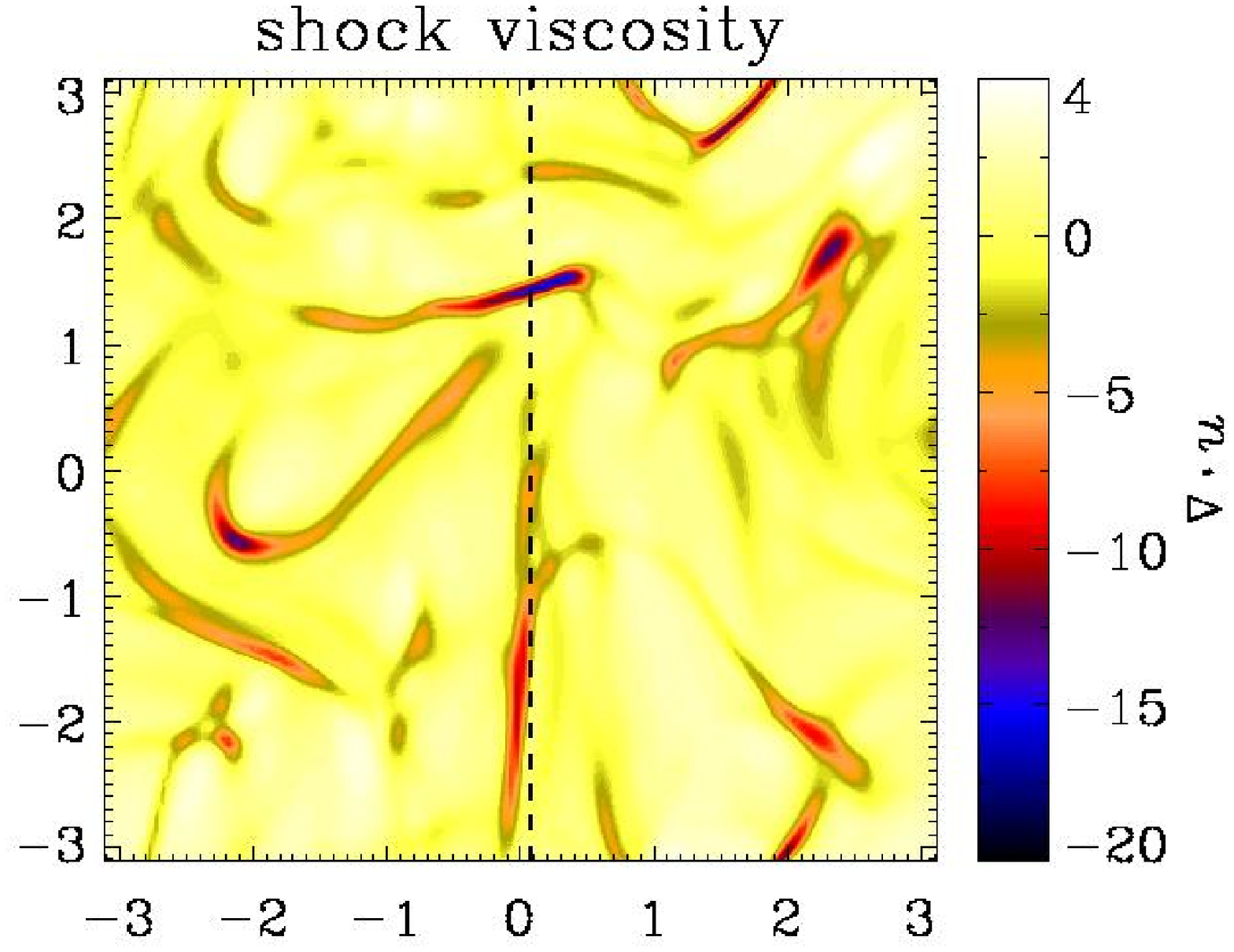}
\caption{
Grey scale (colour scale in electronic version)
representation of $\nab\cdot\uu$ in an $xy$ cross-section
through $z=0$
for $\Ma=1.1$ using constant viscosity (upper panel: Run~3a, 
$512^3$ meshpoints)
and shock-capturing viscosity (lower panel: Run~3b, $128^3$ meshpoints).
Both simulations have been run from an initial state with zero velocity, 
to a state where the kinetic energy has saturated, 
but the magnetic energy is still in the linear regime
(i.e. there is no back reaction on the velocity).
Note the drastic difference between the two regarding the sharpness of
regions of strong convergence.
In both panels, the dotted line indicates the position 
of the scan shown in \Fig{divu_line}.
}\label{vis_divu}\end{figure}

\begin{figure}\centering
\includegraphics[width=0.45\textwidth]{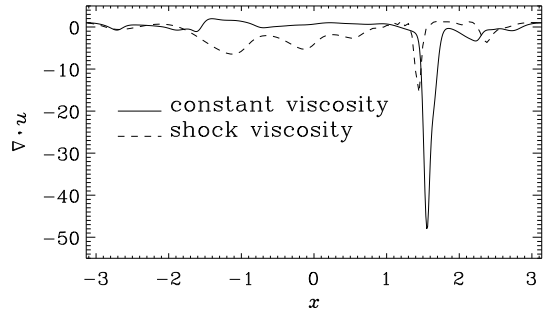}
\caption{
Value of $\nab\cdot\uu$ along a line through the box
whose position is indicated in \Fig{vis_divu}.
Here $\Ma=1.1$ and we are comparing constant viscosity (solid line) 
and shock-capturing viscosity (dashed line).
}\label{divu_line}\end{figure}

\begin{figure}\centering
\includegraphics[width=0.23\textwidth]{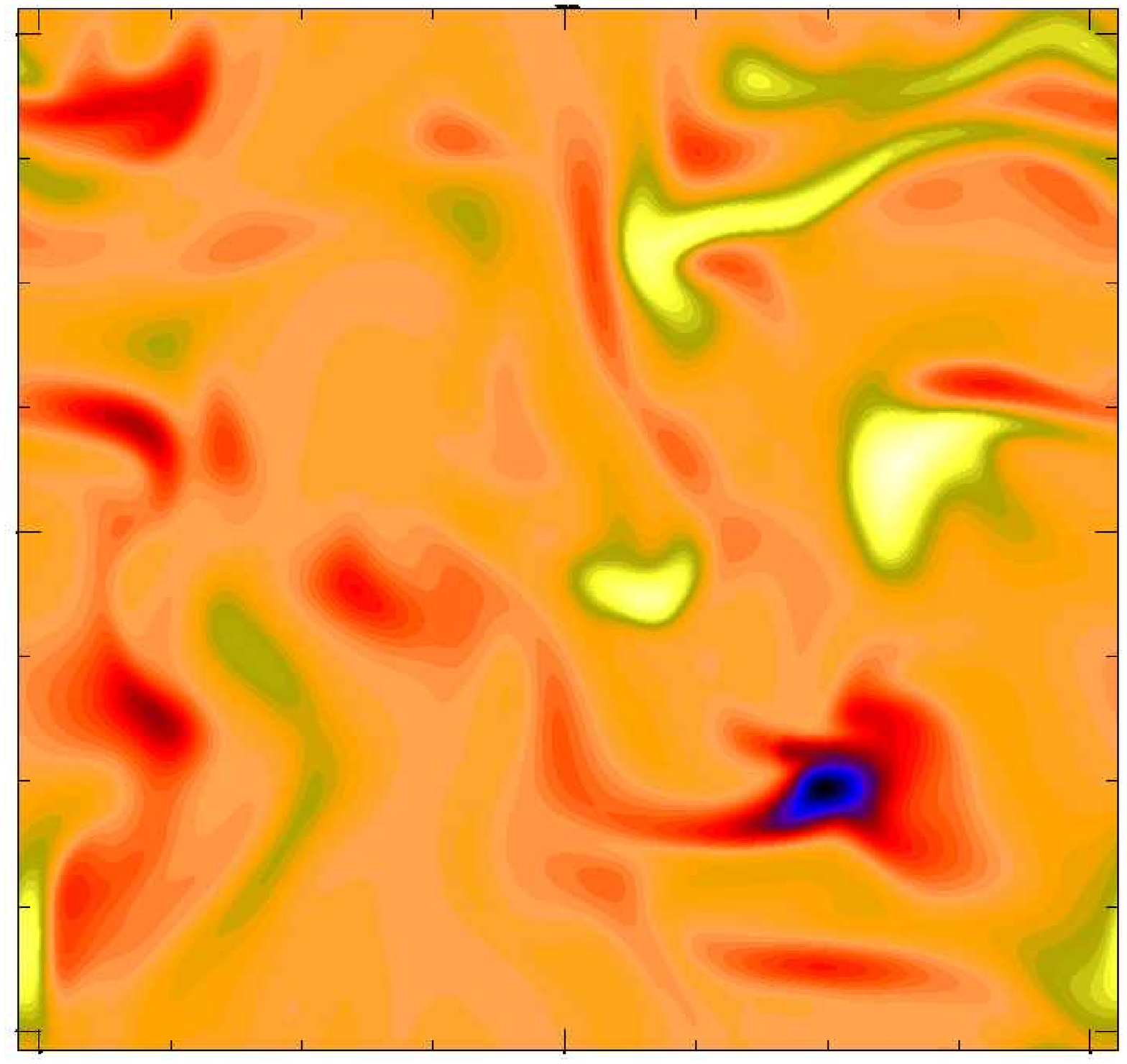}
\includegraphics[width=0.23\textwidth]{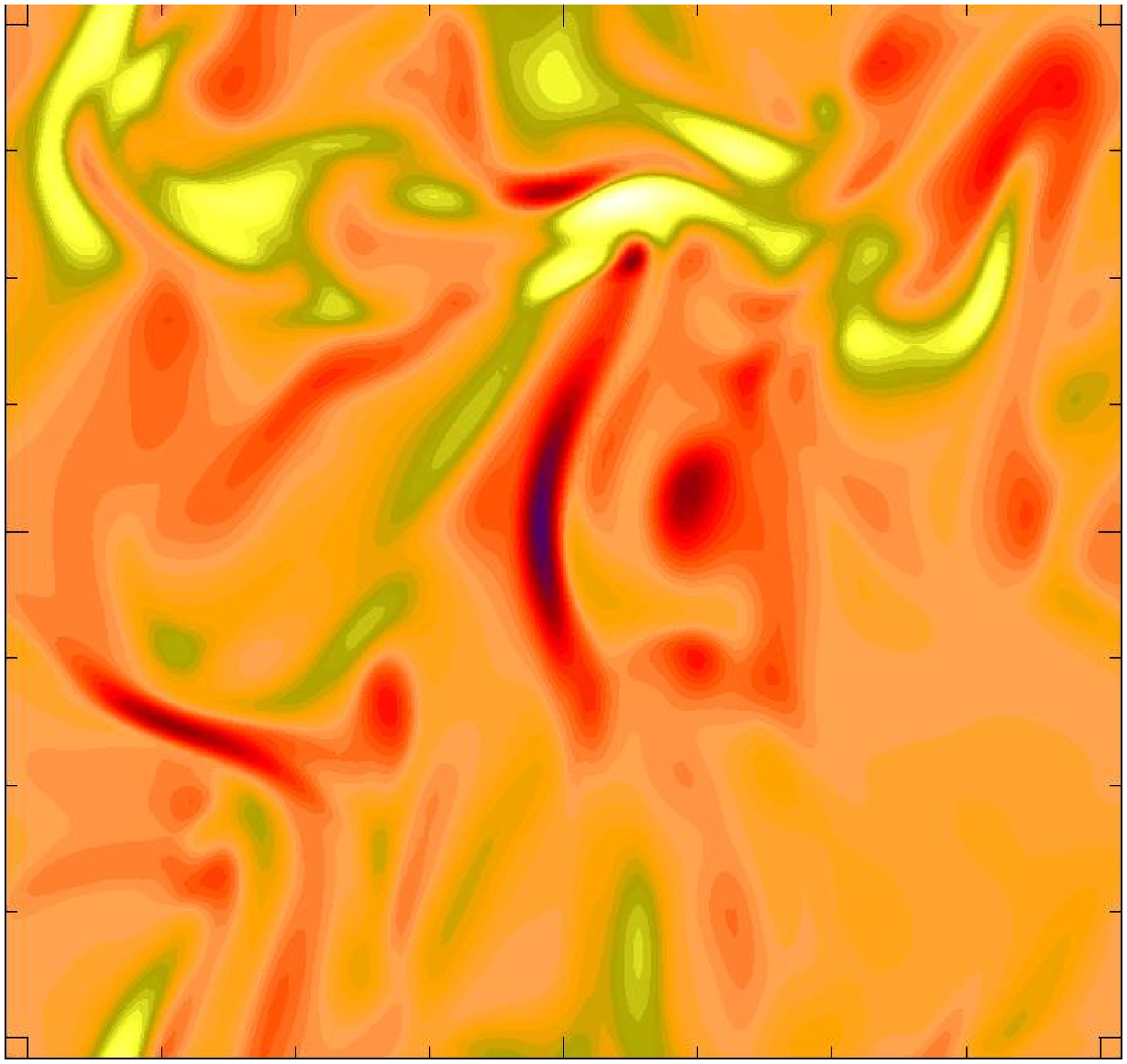}
\caption{
Same as \Fig{vis_divu}, but for $\BB_z$. 
Left figure is from a run with $512^3$ meshpoints and normal viscosity while
right figure is from a run with $128^3$ meshpoints and shock viscosity. 
Since this is in the 
linear regime we have normalized by the rms value.
There is surprisingly little difference between the two.
}\label{vis_bb}\end{figure}

\begin{figure}\centering\includegraphics[width=0.45\textwidth]
{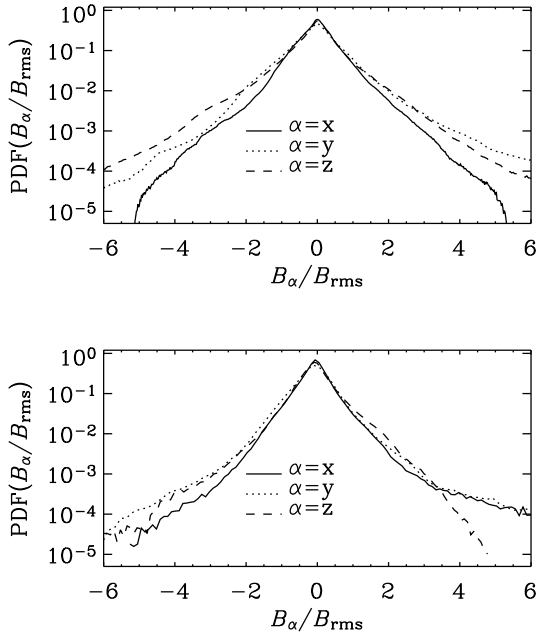}\caption{
Probability distribution function of $B_x$, $B_y$ and $B_z$, normalized
by $B_{\rm rms}$, for Run~3a (upper panel) and Run~3b (lower panel).
}\label{pdf}\end{figure}

\begin{figure}\centering\includegraphics[width=0.45\textwidth]
{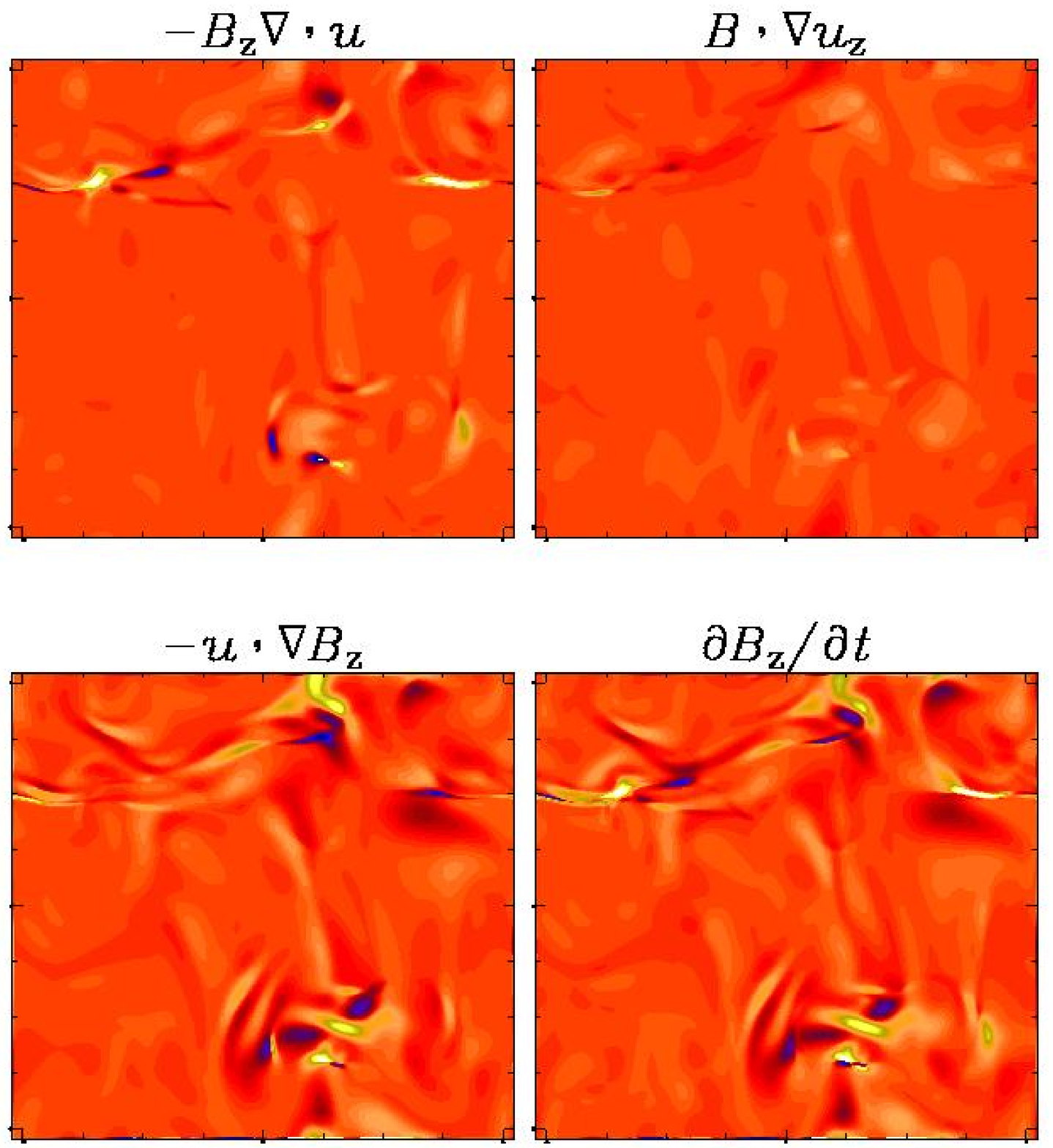}\caption{
The different terms in the induction equation for Run~3a
(with $512^3$ mesh points and constant viscosity). We clearly see that
the dominant term is the advection term.
}\label{induction_tot}\end{figure}

We make a more detailed comparison of two equivalent simulations, one with
and one without shock viscosity.
In each of the two
cases the Mach number is the same, e.g.\ $\Ma\approx1.1$, $f_0=0.5$.
Corresponding cross-sections of $\nab\cdot\uu$ are shown in \Fig{vis_divu}.
In the direct simulations the regions with $\nab\cdot\uu<0$ are quite sharp,
compared with the shock-capturing simulations where they are smoother.
This is seen more clearly in \Fig{divu_line} where we plot
$\nab\cdot\uu$ along a cross-section (shown as a
dashed line in \Fig{vis_divu}).
Since the runs are turbulent, however, direct comparison between individual
structures is meaningless and one can only make qualitative comparisons.

The number of strong convergence regions and shocks is roughly similar
in both cases, but this is only because we have used the same grey/colour
scale in both plots.
In the direct simulation, the dynamical range is
$-62\leq\nab\cdot\uu\leq4.2$, which exceeds the range of the grey/colour scale,
whereas in the shock-capturing simulation the dynamical range is only
$-17\leq\nab\cdot\uu\leq4.2$.
If one were to compare the two cases such that in each the grey/colour scale
is exactly within the dynamical range of that simulation, one would see
more structures in the shock-capturing simulation.

The dynamo-generated magnetic fields
are generally rather smooth in both cases and in that respect
rather similar to each other; see \Fig{vis_bb}.
Indeed, the filling factors where the field exceeds its rms value
are similar in the two cases (0.09 in the left hand panel
for the direct simulation
simulation and 0.13 in the right hand panel for the shock capturing 
simulation).
The probability density functions of the three components of the
magnetic field are in both cases stretched exponentials (\Fig{pdf}), which is
in agreement with earlier results \citep{BJNRST96}.

In order to elucidate the reason for the almost complete absence of
any shock-like structures in the dynamo-generated magnetic field,
we show in \Fig{induction_tot} the different contributions to the
right hand side of $\partial B_z/\partial t$ for Run~3a with
normal viscosity.
The grey/colour scale is the same in all panels.
It is evident that the
dominant term in the induction equation is the advection 
term, $-\uu\cdot\nab B_z$.
The stretching term, $\BB\cdot\nab u_z$, gives the weakest
contribution.
The compression term, $-B_z\nab\cdot\uu$, gives an intermediate
contribution, but, more importantly, there are only a few barely
noticeable contributions from shocks.
This is mainly because in the locations of the shocks
(for example at $x\approx-1.5$ $y \approx 2$; see upper panel
of \Fig{vis_divu}) the normal component of the magnetic field is weak
(\Fig{vis_bb}), making the shocks less pronounced
in the product of the two (i.e.\ in $B_z\nab\cdot\uu$).
The end result is a relatively weak contribution to
the right hand side of $\partial B_z/\partial t$.

\begin{figure}\centering
\includegraphics[width=0.23\textwidth]{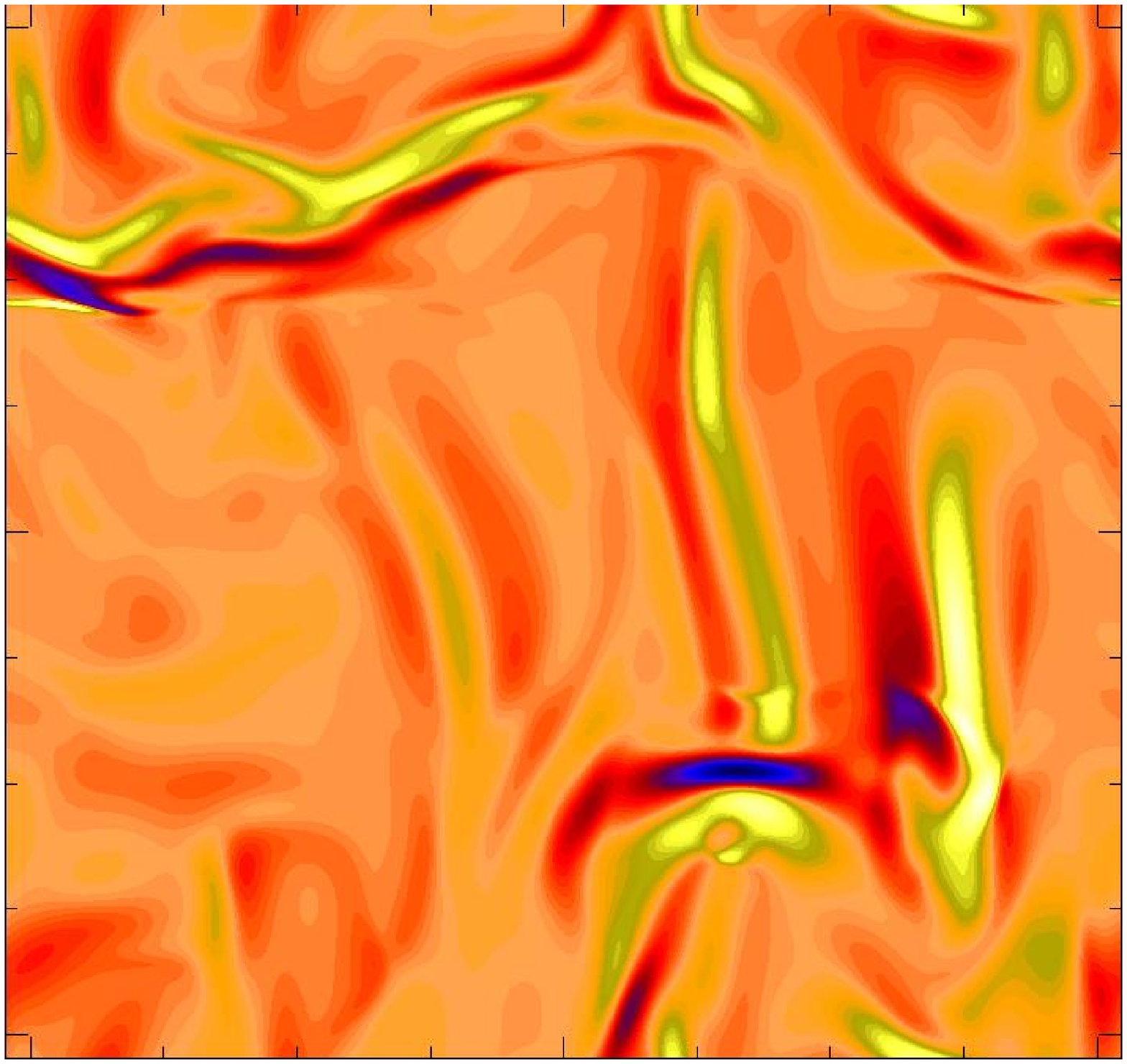}
\includegraphics[width=0.23\textwidth]{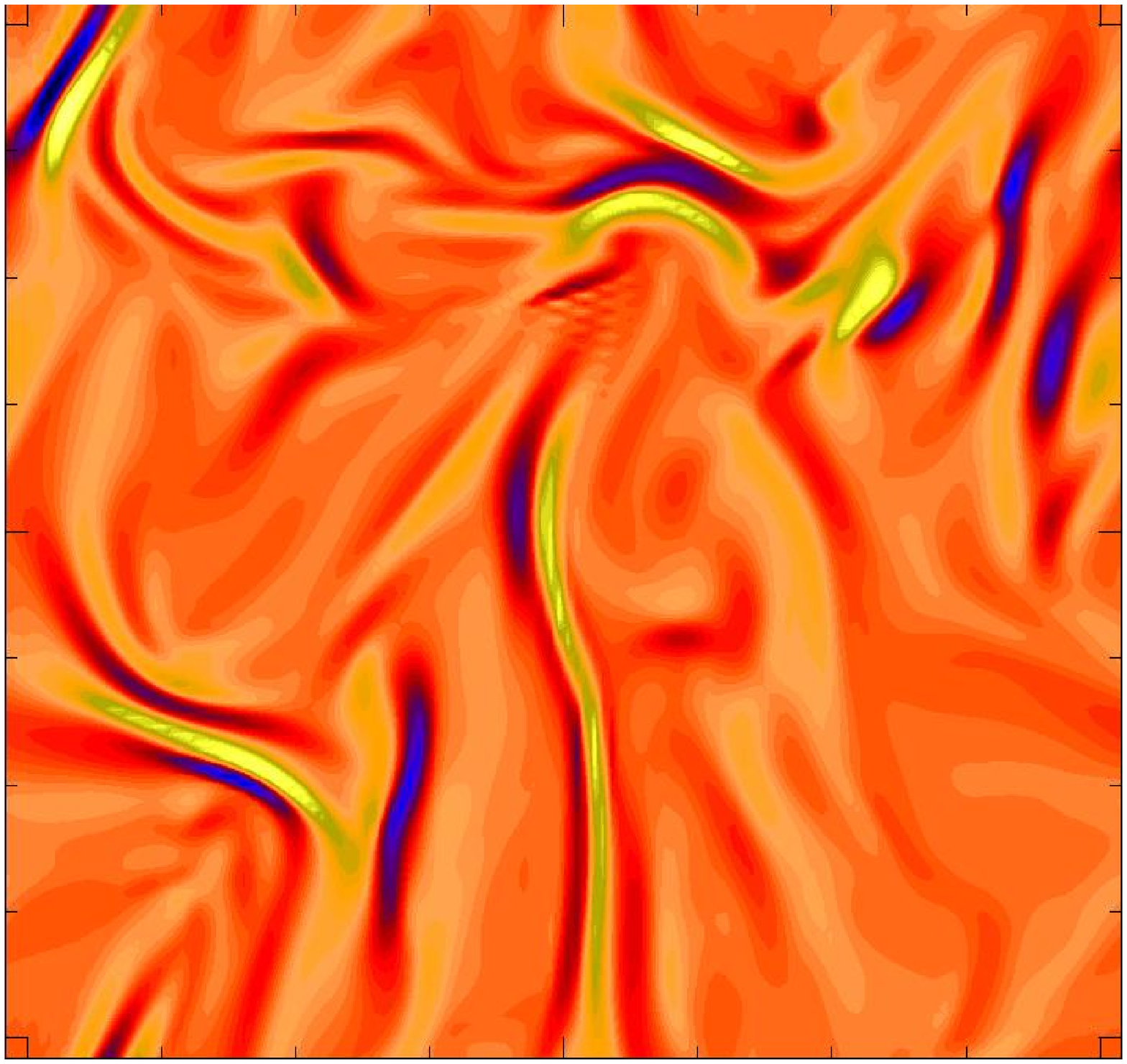}
\caption{
Same as \Fig{vis_bb}, but for $\JJ_z$. Since this is in the 
linear regime we have normalized by the rms value.
The current sheets are similar in thickness, but there are more
of them in the shock-capturing simulation.
}\label{vis_jj}\end{figure}

In the cross-sections of the current density
we see very similar structures in the simulations with normal and 
shock capturing viscosities; see \Fig{vis_jj}. This is supported by the
filling factors which are $\approx 0.1$ for both the left and the right 
hand sides of the figure.

\begin{figure}\centering\includegraphics[width=0.45\textwidth]
{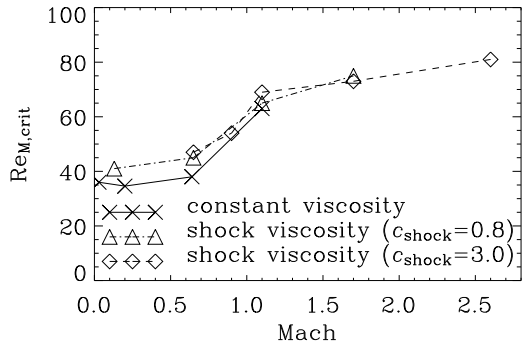}\caption{
Critical magnetic Reynolds number $\Rey_{\rm M,crit}$ as a function of $\Ma$
for simulations with $\Pm=1$.
Note that $\Rey_{\rm M,crit}$ depends strongly on Mach number for
$\Ma\approx1$.
The simulations with shock-capturing viscosity give approximately
the correct growth rates. The simulations that provide these data points
have resolutions ranging from $64^3$ to $512^3$ mesh points.
(Some of them are listed in \Tab{Trunscompare}.)
}\label{Mach_dep_Rmcrit}\end{figure}

\begin{figure}\centering\includegraphics[width=0.45\textwidth]
{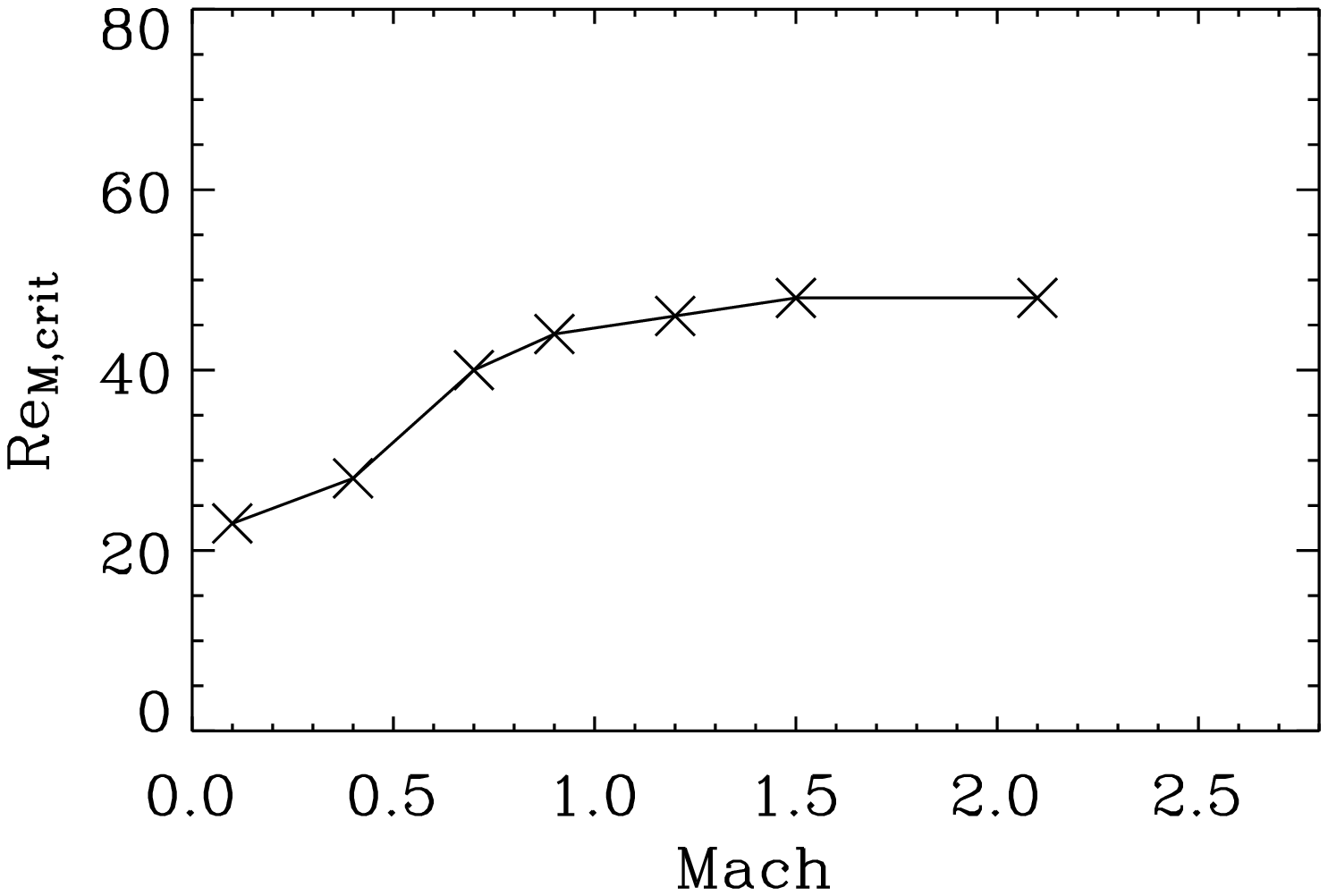}\caption{
Same as \Fig{Mach_dep_Rmcrit} but for simulations with $\Pm=5$.
}\label{Mach_dep_Rmcrit_Pm5}\end{figure}

In \Fig{Mach_dep_Rmcrit} we plot the critical magnetic Reynolds number
$\Rey_{\rm M,crit}$ as a function of Mach number.
It turns out that the critical magnetic Reynolds number increases from
about 35 in the subsonic regime to about 70 in the supersonic regime.
Whether or not the critical magnetic Reynolds number increases even
further for larger Mach numbers is unclear, because larger resolution
is required to settle this question for $\Pm=1$.
However, in the direct simulations with $\Pm=5$
(\Fig{Mach_dep_Rmcrit_Pm5}), and in the shock-capturing simulations with
$\Pm=1$ the critical magnetic Reynolds number is roughly unchanged when
$\Ma$ is increased beyond $\Ma\approx1$.
We refer to this possibility of having two distinct values of the
critical magnetic Reynolds number for subsonic and supersonic turbulence, with
a reasonably sharp transition at $\Ma\approx1$, as `bimodal' behaviour.

Comparing direct and shock-capturing simulations we find the general
appearance of the cross-sections of $\BB$ remarkably similar.
The simulations with shock-capturing viscosity have, however, slightly larger
critical magnetic Reynolds numbers.
This is probably explained by the smaller velocity gradients and hence smaller
stretching rates in the simulations with additional shock-capturing
viscosity.
Nevertheless, both direct and shock-capturing simulations show a
roughly similar functional dependence of the critical magnetic Reynolds
number on the Mach number.
This suggests that shock-capturing simulations provide a reasonable
approximation to the much more expensive direct simulations--at least
as far as the onset of turbulent dynamo action is concerned.

Once the Mach number becomes comparable to unity, one needs very high
resolution to resolve the rather thin shock structures.
However, as seen above, the magnetic field structures are by far not
as thin as the hydrodynamic shocks.
This implies that one can decrease $\eta$ to values far below $\nu$
before the magnetic field and current structures would become unresolved.
In other words, with given resolution, one can reach a magnetic Reynolds
number that is much larger than the kinematic Reynolds number.
Alternatively, if one only wants to reach just weakly supercritical
values of $\Rm$ for dynamo action (which turn out to be roughly below
one hundred), $\Rey$ can be much smaller and just a few tens.
This regime corresponds to large values of $\Pm$.
For $\Pm=5$, for example, we are able to reach Mach numbers of up to 2.1.
The resulting plot of $\Rey_{\rm M,crit}$ versus $\Ma$
is shown in \Fig{Mach_dep_Rmcrit_Pm5}.
This plot confirms again the approximately `bimodal' behaviour of
the critical magnetic Reynolds number, but now
$\Rey_{\rm M,crit}\approx25$ in the subsonic regime and
$\Rey_{\rm M,crit}\approx50$ in the supersonic regime.

\begin{figure}\centering\includegraphics[width=0.45\textwidth]
{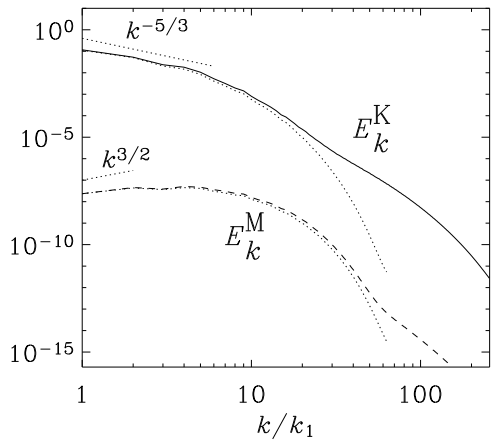}\caption{
Energy spectra for Runs~1a and 1c.
The dotted lines give the result using shock-capturing viscosity
(numbers in parentheses in the table).
}\label{spectra}\end{figure}

The kinetic and magnetic energy spectra show close agreement between
the constant and shock-capturing viscosity solutions at low wavenumbers;
see \Fig{spectra}.
The truncation at higher wavenumbers is presumably a direct consequence of
the locally enhanced shock viscosity which attempts to
increase the width of the shocks.
Furthermore, since shocks hardly manifest themselves in the magnetic
field, the agreement between the magnetic energy spectra in the direct
and shock-capturing simulations extends to even larger wavenumbers.
As discussed in the beginning of \Sec{SResults}, the direct simulations
require vastly more resolution, but once the resolution is sufficient,
either of the two spectra agrees well when doubling the resolution;
see also the middle panel of Fig.~13 in \cite{HBD04}.

\section{Discussion}

The present results have shown that in both direct and shock-capturing 
simulations the onset of dynamo action requires a somewhat 
larger magnetic Reynolds number when the Mach number exceeds a critical 
value around unity.
In the subsonic regime the critical magnetic Reynolds number for
nonhelical dynamo action is around 35, but it increases to about
70 in the supersonic regime.
Once the Mach number exceeds unity, the critical magnetic
Reynolds number no longer seems to depend on the Mach number.
This confirms the notion that in the supersonic regime
dynamos experience an additional sink.
This additional sink is plausibly related to the sweeping up of magnetic
field by shocks, but this effect does not gain in its importance with
increasing Mach number once we are already in the supersonic regime.

\begin{figure}\centering\includegraphics[width=0.40\textwidth]
{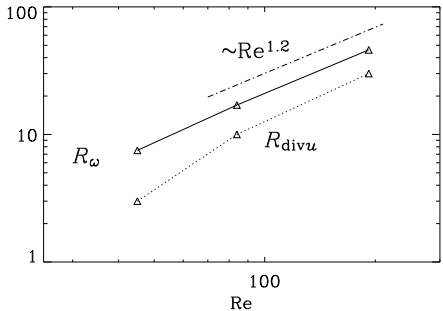}\caption{
Dependence of $R_{\dive\uu}$ and $R_\omega$ on $\Rey$.
Note that both quantities remain well separated as they
tend to approach a slope of 1.2.
}\label{pRe_omega_divu}\end{figure}

The relative importance of the adverse effects of shocks can be
associated with a Reynolds number-like quantity
\EQ
R_{\dive\uu}=\left.
\bra{(\nab\cdot\uu)^2}^{1/2}\right/\left(\nu k_{\rm f}^2\right),
\EN
whose dependence on the actual Reynolds number is shown in
\Fig{pRe_omega_divu}.
Here we also compare with a similar number quantifying the
relative importance of vortical motions,
\EQ
R_\omega=\left.
\bra{\oo^2}^{1/2}\right/\left(\nu k_{\rm f}^2\right).
\EN
Both numbers seem to approach a limiting behaviour proportional to
$\Rey^{1.2}$.
On theoretical grounds, one would have expected a slope of 3/2,
because the ratio of rms velocity to
rms vorticity is proportional to the Taylor microscale which, in turn,
is known to diminish proportional to $\Rey^{-1/2}$.
The departure from this expectation might indicate that we are not
yet in the asymptotic regime with a well developed inertial range.
This is also plausible from \Fig{spectra} showing that most of
the spectrum is dominated by a rather extended dissipative subrange
with only a rather short inertial range.
Thus, it might not be surprising that the Taylor microscale does not
yet show a clear asymptotic $\Rey^{-1/2}$ scaling.

\FFig{pRe_omega_divu} suggests that $R_{\dive\uu}$ remains smaller than
$R_\omega$ by a constant factor of around 1.5.
Thus, $\bra{(\nab\cdot\uu)^2}/\bra{\oo^2}\approx0.44$.
This ratio would be exactly 1/2 if the mean square values of
longitudinal and transversal velocity derivatives were equal, i.e.\
$\bra{u_{x,x}^2}=\bra{u_{x,y}^2}$.
Here we have assumed isotropy and that mixed terms cancel, which implies
$\bra{(\nab\cdot\uu)^2}\approx3\bra{u_{x,x}^2}$ and
$\bra{\oo^2}\approx6\bra{u_{x,y}^2}$, giving a ratio of $1/2$.

Obviously, it will be important to confirm the asymptotic behaviour
of $R_\omega^2/R_{\dive\uu}^2$ at larger
Reynolds and Mach numbers and to relate this to the value of the critical
magnetic Reynolds number for dynamo action.
At the moment our simulations are simply limited by the resolution
($512^3$ meshpoints) which in turn is limited by the computing power
available on modestly big supercomputers.
Even though somewhat larger resolutions are already possible ($1024^3$
meshpoints), those runs remain prohibitively expensive if one needs to
run for many turnover times.

Although we have here focused on the critical value of the magnetic
Reynolds number for dynamo action, rather than the growth rate, we can
see that there is currently no evidence for an increasing growth rate
with increasing Mach number.
From Runs~1a and 3a in Table~\ref{Trunscompare},
one sees that,
as $\Ma$ increases from 0.72 to 1.1, the growth rate actually decreases,
even though the magnetic Reynolds number is the same.
By comparison, early work on small scale dynamo action in
irrotational turbulence \citep{KRS85} has shown that the growth
rate should increase with the Mach number to the fourth power.
Although we cannot confirm this, it is important to bear in mind
that irrotational turbulence is not likely to correspond to the limit
of large Mach numbers.
Instead, as discussed above, the irrotational contribution to the
flow velocity may converge to a finite fraction of the total flow
velocity in the limit of large Mach numbers.

\section{Conclusions}

The present simulations suggest that the critical magnetic Reynolds
number for the onset of dynamo action switches from about 35 in the
subsonic regime to about 70 in the supersonic regime, and stays then
approximately constant (at least in the range $1.0\leq\Ma\leq2.6$ that
we were able to simulate).
It turns out that the simulations with shock-capturing viscosity
yield almost the same critical magnetic Reynolds number as the
direct simulations.
This is partly explained by the fact that the shock structures are
approximately equally smooth for both direct and shock-capturing
viscosities.
Finally, we note that there is as yet no support for the
expectation that the growth rate is proportional to the Mach number
to the fourth power, as was suggested previously by analytic theory
assuming that the flow is irrotational \citep{KRS85,MS96}.
It is suggested that this is because supersonic turbulence cannot
accurately be described as being fully irrotational.
Indeed, we find that the Reynolds
number based on the vortical component of the flow is always larger than
the Reynolds number based on the irrotational component.

In order to verify the proposed asymptotic independence of the critical
magnetic Reynolds number on the Mach number, it would be useful to extend
both the direct and the shock-capturing simulations to larger values of
the Mach number.
An obvious reason why this may not have been done yet is that when shock
capturing viscosities are used, one normally uses at the same time also
a similarly defined shock-capturing magnetic diffusivity.
Consequently, a definition of the usual magnetic Reynolds number is no longer possible,
and the connection between effective magnetic Reynolds number (based on
the shock-capturing magnetic diffusivity) and the actual one is unclear.
This is why in the present work we kept the magnetic diffusivity and the
kinematic viscosity equal to the microscopic one.
Even when we compared with shock-capturing simulations,
the magnetic diffusivity was still kept constant.
In conclusion, we feel that the use of shock-capturing viscosities
in dynamo simulations with constant magnetic diffusivities provides a
reasonable tool for investigating supersonic hydromagnetic turbulence.

\section*{Acknowledgements}
We are indebted to Anvar Shukurov for making useful suggestions
regarding the discussion of irrotational flows.
We also thank {\AA}ke Nordlund and Ulf Torkelsson for suggestions,
corrections and comments on the paper.
We acknowledge the Danish Center for Scientific Computing
for granting time on the Horseshoe cluster,
and the Norwegian High Performance Computing Consortium (NOTUR)
for granting time on the parallel computers in
Trondheim (Gridur/Embla) and Bergen (Fire).  Travel between Newcastle 
and NORDITA was also supported by a travel grant from the Particle 
Physics and Astronomy Research Council.


%
\label{lastpage}
\end{document}